# Compression-induced magnetic obstructed atomic insulator and spin singlet state in antiferromagnetic KV$_2$Se$_2$O

Liucheng Chen[1†], Jiayi Yue[1,2†], Jingwen Cheng[1,2†], Jianli Bai[1,2], Zexiao Zhang[4,1], Xiaoli Ma[1], Fang Hong[1,2,3], Genfu Chen[1,2,3], Jian-Tao Wang[1,2,3*] Zhijun Wang[1,5*] & Xiaohui Yu[1,2,3*]

**Affiliations:**

[1]*Beijing National Laboratory for Condensed Matter Physics and Institute of Physics, Chinese Academy of Sciences, Beijing, China*

[2]*School of Physical Sciences, University of Chinese Academy of Sciences, Beijing, China*

[3]*Songshan Lake Materials Laboratory, Dongguan, Guangdong, China*

[4]*Department of Physics, Chengdu University of Technology, Chengdu, China*

[5]*Condensed Matter Physics Data Center, Chinese Academy of Sciences, Beijing 100190, China*

†These authors contributed equally to this work.

**Abstract: Among the complex many-body systems, the metal-insulator transition stands out as a cornerstone and a particularly fertile ground for scientific inquiry [1-3]. The established models including Mott insulator, Anderson localization and Peierls transition, are still insufficient to capture the complex and intertwined phenomena observed in certain material systems [4-10]. KV$_2$Se$_2$O, a newly discovered room-temperature altermagnetic candidate exhibiting a spin-density-wave transition below 100 K, provides a unique platform to investigate the interplay of many-body effects and unconventional magnetism, specifically the anticipated metal-insulator transition under extreme conditions. Here, we report a compression-induced insulator by suppressing the metallic**

**behavior without structural phase transition. The newly opened gap is estimated to be ~ 40 meV at around 43.5 GPa, given direct evidence for the insulating state. A concurrent switching of carrier type demonstrates the large Fermi surface reconstruction crossing the metal-insulator transition. The density functional theory calculations indicate that the discovered $V^{+2.5}$-based insulator is a magnetic obstructed atomic insulator, being a spin-singlet state with bonding orbital order. This work not only presents an archetype of a pressure-driven metal-insulator transition decoupled from structural change but also delivers fundamental physical insights into the metal-insulator transition.**

**Introduction**

Metal-to-insulator (MI) transition represents one of the most fundamental and enduring challenges in condensed matter physics [1-3]. Historically, the challenge of explaining this phenomenon directly spurred the development of seminal theoretical frameworks, most notably Peierls' theory of symmetry breaking (thermodynamic phase transition) [4-6], Mott's theory of strong electron interactions and Anderson's theory of localization [7-10]. Generally, the MI transition is rarely an isolated event, which is intimately linked to a host of other emergent quantum phenomena, including high-temperature superconductivity and colossal magnetoresistance [11-14]. It is these complex and intertwined phenomena that reveal a gap between existing theories and the mechanisms of certain MI transitions [1-3]. The MI transition is sensitive to stimuli including temperature, pressure, composition, and many other means [11-17]. Among them, high pressure is an important and effective tool for creating a new ground state of materials through tuning of crystal structures and electronic structures [18-20]. The pressure-induced metallization of insulators is a common phenomenon, because of the band broadening and gap closure based on the band theory [21-23]. In contrast, pressure-driven metal to insulator transition is comparatively rare and usually accompanied by first-order structural transitions [24-26]. Therefore, the investigation of pressure induced MI transition without structural transition enables a unique strategy for elucidating the underlying physical mechanisms of the insulating state for the complex and intertwined materials.

Very recently, a room-temperature altermagnetic candidate $KV_2Se_2O$ with *d*-wave spin-momentum locking confirmed by the spin-and angle-resolved photoemission spectroscopy measurements have been reported [27,28]. The magnetic moments of the V atoms form long-range order but with zero net magnetization in the measured temperature range. Transport and magnetic measurements reveal a density wave (DW) -like transition in this compound with a critical temperature ~ 105 K [29,30]. The results from nuclear magnetic resonance and angle-resolved photoemission spectroscopy indicate the spin density wave (SDW) transition below 100 K with an in-plane period of $\sqrt{2}a \times \sqrt{2}a$ and the preservation of altermagnetic order in the SDW phase [28]. The emergence of SDW order in $KV_2Se_2O$ provides a unique opportunity for explore intriguing quantum phenomena such as unconventional superconductivity as observed in the cuprates and iron-based superconductors [31,32], and even in intensely studied nickelates in latest a few years [33]. The strong Coulomb interaction between the 3*d* electrons of vanadium atoms make vanadium sesquioxide and its derivatives as the canonical Mott-Hubbard system [1,13,15,26]. In addition, the behavior of this material under pressure is yet to be characterized. Thus, the altermagnetic candidate $KV_2Se_2O$ is a unique platform to investigate the interplay of many-body effects coupled with unconventional magnetism, particularly the expecting superconductivity and MI transition under extreme conditions.

To investigate the complex interplay under extreme conditions, we performed high-pressure studies on the altermagnetic candidate $KV_2Se_2O$ up to 60 GPa using a diamond-anvil cell. High-pressure X-ray diffraction (XRD) revealed a structural transition from *P4/mmm* to *Pmmn*, with a narrow coexistence region under pressure. Resistivity measurements showed that the SDW order is gradually suppressed and vanishes near 10 GPa. Above 35 GPa, a dome-like resistivity behavior emerges, signaling an insulating state in the *Pmmn* phase. The extracted band gap provides direct evidence of a MI transition. The observed switching of charge carriers provides strong evidence for Fermi surface reconstruction as the underlying driver. Band structure calculations further indicate that the transition originates from a magnetic obstructed atomic insulator.

**Results and discussion**

**Characterizations at ambient pressure:** Before the measurements under pressure, crystal structure and resistivity of $KV_2Se_2O$ at ambient pressure were measured to check the crystal quality (Fig. 1). The single-crystal structure of $KV_2Se_2O$ was measured by the single-crystal XRD system with a Mo x-ray source. The refinement gives a tetragonal structure with the space group of $P4/mmm$. The refined lattice parameters at room temperature are $a$ = 3.9698(3) Å, $c$ = 7.3369(8) Å (Table S1), which is almost consisted with previous report [27-30]. Then, the characterization of resistivity $\rho_{xx}$ for the single-crystal sample at ambient pressure was also carried out by using the four-probe method. As observed in previous work [29], the temperature-dependent $\rho_{xx}$ of $KV_2Se_2O$ also display the characteristics of a metallic behavior (Fig. 1f). With the increase of temperature, $\rho_{xx}$ firstly increases sharply, and then saturates with further increasing temperature, after passing a kink at around 105 K. This kink, showing a metal-insulator-metal-like transition, reminiscent of quasi-two-dimensional CDW/SDW materials, was attributed to an SDW transition [28,29].

**Metal-insulator transition under pressure.** Under pressure, the SDW transition is suppressed, and completely suppressed above 9 GPa (Fig. 2b, upper panel). The low-temperature $\rho_{xx}$ in Fig. 2a is fitted by the power law (Supplementary Note I). The suppressed value of $n$ from 3.6 at ambient pressure to 2.3 at 9 GPa (Fig. 2b, middle panel), indicates the interaction of $KV_2Se_2O$ at low temperatures shifting from the strong electron-phonon interaction to a weak correlated metal [34,35]. As displayed in Fig. 2c and Fig. S1, the $\rho_{xx}$ shows insensitivity to the external compression between 20-30 GPa, with only a slight increase observed. Subsequently, the resistivity exhibits a monotonic increase with pressure, developing a hump at low temperatures. Clearly, the hump-like anomaly shifts progressively downward in temperature with increasing pressure. This electrical transport anomaly provides direct evidence for a pressure-driven MI transition in $KV_2Se_2O$. The metal-insulator transition temperature $T_{MI}$ is determined from the valley in $dR/dT$ (Fig. S4). As can be seen from Fig. 2d (upper panel), the insulating state maintains the entire temperature range above 50 GPa. The small deviations of the $T_{MI}$ for the two runs could be caused by the different errors of the two electrical measurement systems.

Then, the band-gap ($\Delta$) of the insulating state was fitted (Supplementary Note II and Fig. S5) and summarized in Fig. 2d (middle panel). The $\Delta$ value increases markedly upon compression, reaching a maximum of ~40 meV at around 43.5 GPa, then decreases monotonically at higher pressures. The opened gap of $KV_2Se_2O$ under pressure gives solid evidence for the pressure-driven MI transition. Above ~50 GPa, although $\rho_{xx}$ subsequently decreases with further compression, the system maintains its insulating character up to the highest pressure studied in this work. Furthermore, the $\rho_{xx}$ at 2 K in run 1 and 4.5 K in run 2 under pressure (Fig. 2d, low panel) are extracted from the temperature-dependent $\rho_{xx}$ (Fig. 2c and Fig. S1b). Under applied pressure, the resistivity $\rho_{xx}$ at 2 K and 4.5 K exhibits an initial gradual increase up to 30 GPa, followed by a steep rise to 40 GPa, before undergoing progressive decline at higher pressures. Obviously, this sharp enhancement of $\rho_{xx}$ correlates directly with the emergence of insulating behavior.

**Structural properties at high pressures.** The high-pressure structural behaviors of $KV_2Se_2O$ were uncovered by the measurements of synchrotron XRD. The XRD patterns and refined results are presented in Fig. 3 and Fig. S6. The initial $P4/mmm$ phase is persistent up to 22 GPa. Above 24 GPa, the changes in the diffraction patterns are completed, which means the second phase maintains the crystal structure of $KV_2Se_2O$. To further gain insights into the lattice evolution with pressure, we refine the XRD patters under pressure by using the Rietveld method [36]. The initial phase was refined with the tetragonal structure with space group of $P4/mmm$, which is the same as that $KV_2Se_2O$ at ambient pressure (Fig. S6 and Supplementary Note III) [29]. Upon compression, the $Pmmn$ phase becomes more stable compared to the $P4/mmm$ phase above 12 GPa, depending on the calculated energetic data (Fig. 3b). Typical refining examples at 25.4, 34.1 and 48.4 GPa with an acceptable error range are shown in Fig. 3c. Figure 3d presents the crystal structure and the magnetic structure of $KV_2Se_2O$. The detailed structure information of this new phase is listed in Table S2. Depending on the refinements, the pressure dependent lattice parameters and volume of $Pmmn$ phase are plotted in Fig. 4e. Over the pressure range of 22 to 58 GPa, the crystal structure of KV$_2$Se$_2$O was found to adopt the *Pmmn* space group, and no further transitions were detected up to 58 GPa. Therefore, the electronic gap observed above 35 GPa cannot be attributed to a structural phase transition, since it occurs at a pressure far beyond the phase-transition point.

**Reconstruction of the Fermi surface.** Magnetoresistance (MR) and Hall coefficient ($R_H$) exhibit pronounced sensitivity to charge-order transitions, serving as a diagnostic indicator for detecting scattering anisotropy changes and Fermi surface reconstruction [37,38]. The MR measured at various pressures and the analyses are shown in Supplementary Note IV and Fig. S7. In the initial phase, the MR at 8 T has an enormous increase together with the suppression of SDW order at low pressures (Fig. 4a). The tremendous MR at low pressures indicates a strong coupling between the SDW state and the charge carriers. Then, the MR monotonously decreases until to near 20 GPa and reserves a negligible value. Traversing a narrow pressure window, the value of MR at 8 T progressively decreases with further compression up to 40 GPa (Insert of Fig. 4a). Concurrently, the band-gap $\Delta$ progressively increases to its maximum value (Fig. 3d). Above 40 GPa, coincident with a slight enhancement of MR, $\Delta$ undergoes continuous reduction up to 56 GPa. Depending on the Hall resistivity measured at 2 K under various pressures (Supplementary Note V and Fig. S8), we can obtain the $R_H$ and $n_H$. As shown in Fig. 4b-4c, two unusual kinks can be clearly seen at around 15 and 40 GPa. The first kink reflects the emergency of the *Pmmn* phase (Fig. 3). The climbing up $n_H$ supports the decreased $\rho_{xx}$ between 12 and 20 GPa (Fig. 2). Surprisingly, at the second unusual kink, the carrier type switches from hole type to electronic type. The switched carrier type indicates a large reconstruction of the Fermi surface. In addition, the suppressed $n_H$ from ~$10^{22}$ cm$^{-3}$ to $10^{20}$ cm$^{-3}$ at higher pressures is consistent with the observed insulating state in the *Pmmn* phase. Consequently, the modification of electronic states near the Fermi surface and the decreased carrier concentrations are expected to be the dominant factor driving the insulator phase.

**Band structure calculations.** As displayed in the temperature-pressure (T-P) phase diagram of KV$_2$Se$_2$O in Fig. 4d, The *Pmmn* phase initially exhibits metallic behavior below approximately

35 GPa. It then undergoes a clear MI transition upon further compression, with the insulating state becoming the dominant configuration beyond 45 GPa. The theoretical calculations confirm this experimental finding. In detail, the *Pmmn* phase originates from the *P4/mmm* phase through V-V dimerization mediated by O atoms, which results in two distinct V-V distances (i.e., the short one ~3.36 Å, and the longer one ~3.82Å). By wiggling the V-O-V chains, the lattice constants decrease under pressure. The V-O-V bond angles change from 180 degrees of the *P4/mmm* phase to 116 degrees of the *Pmmn* phase (Fig. S10). The V-V dimerizations along the *x* and *y* directions intersect at the O atoms, which occupy the 2b Wyckoff positions. The absence of imaginary frequencies in the phonon spectrum under pressure indicates the structural stability of the *Pmmn* phase (Fig. 5b). At the initial of *Pmmn* phase (Fig. S10) under pressure, the electronic band structure shows a typical metallic character, which is consistent with the experimental results (Fig. 2).

Then, the electronic band structures and density of states (DOS) were computed at 43.5 GPa. A comparison of the band structures with and without spin-orbit coupling (SOC) reveals that SOC has a minor impact on this system (Fig. S11). Therefore, we focus on the local spin density approximation (LSDA)+U calculations in the absence of SOC in the main text. Several collinear magnetic configurations were investigated (Fig. S12). Among them, the configuration depicted in Fig. 5a represents the most energetically favorable state. Its LSDA+U band structure is shown in Fig. 5d, which exhibits a direct band gap of about 0.1 eV. The insulating behavior is consistent with the resistivity measurements shown in Fig. 2. Because no symmetry operation connects the spin-up $V_{4f}$ atoms and the spin-down $V_{4e}$ atoms, the DOS of the two spin channels in Fig. 5c is asymmetric. Additionally, the valence band maximum (VBM) and conduction band minimum (CBM) are mainly contributed by $V_{4f}$-$d$ and $V_{4e}$-$d$ electrons, respectively. From atomic valence states, we conjecture that the V atoms are of the $V^{+2.5}$ valence state, which has the $d^{2.5}$ configuration. The insulating behavior of $d^{2.5}$ V-based materials has not been reported before.

In order to analyze the orbitals of this $V^{+2.5}$-based compound, we focus on the highest 46 valence bands in the spin-up channel (from -9 to 0 eV), which are well separated from other bands (Fig. 5c). The atomic-valence electron band representations (ABRs) are summarized in Table S3. Based on the computed irreducible representations (irreps) obtained using the IRVSP program [39], the ABRs of these bands are decomposed [40] into $A@8g(Se-p) + (A1+B2+B1)@2a(O-p) + (A1+B2+B1)@2b(O-p) + (A'+A'')@4f(two\ V_{4f}-d) + A1@2b\ (bonding\ d_{xz})$ (Table S3). Therefore, we conclude that the fully occupied Se-$p$ states are from -6 to -2 eV, the fully occupied O-$p$ states are from -9 to -6 eV, while some V-orbitals are slightly below $E_F$. These results align well with the characteristics of fatted-band structures (Se-$p$, O-$p$, and $V_{4f/4e}$-$d$ orbitals) of Fig. S9 in the supplementary materials. The ABR decomposition shows that it is unconventional with an essential A1@2b elementary band representation (EBR), being an obstructed atomic insulator formed by the V-$d$ orbitals. On the other hand, the ABR decomposition of the spin-down bands shows that the two $V_{4e}$-$d$ orbitals and one bonding $d_{yz}$ orbital are occupied. Thus, the spin-up bonding $V_{4f}$-$d_{xz}$ and the spin-down bonding $V_{4e}$-$d_{yz}$ orbital actually form a spin-singlet state with bonding orbital order. Our calculations show that both Hubbard U and the magnetic configuration are crucial to obtain the insulating band structure. This spin singlet state is unique with a special type of spin-orbit coupling, the spin-up state coupled with bonding V-$d_{xz}$ orbital and the spin-down state coupled with bonding V-$d_{yz}$ orbital. This state resembles the ground state of the Heisenberg spin model ($J\ \hat{S}_1 \cdot \hat{S}_2$) with ferromagnetic coupling ($J > 0$). Additionally, the unconventional nature of the OAI is characterized by the RSI $\delta_{2b} = -1$. Thus, the obstructed surface states are expected on the side surfaces. By constructing the Wannier tight-binding Hamiltonian, the 100-surface spectrum is obtained in Fig. 5f with the slab calculation. The obstructed surface states are clearly shown as the blue-colored and red-colored on the left and right surfaces, consistent with the unconventional OAI nature.

To understand the formation of the insulating behavior of the $V^{+2.5}$ state, we draw the schematic diagram (Fig. 5e). Under $C_{4v}$ symmetry, the fivefold $d$ orbitals split into doubly degenerate $d_{xz,yz}$, doubly degenerate $d_{x2-y2,xy}$ states and the $d_{z2}$ state. Since there is no apical coordinating atom, the $d_{z2}$ state is lowest. Under the $C_{2v}$ (2mm) site symmetry of V atom and the corresponding crystal field of the compound, the energy levels of $d_{yz}$ and $d_{x2-y2}$ states are increased, due to the hybridization with the lower Se-$p$ states and the $d_{z2}$ states, respectively. Among the three lower states, $d_{z2}$ is the lowest, while $d_{xz}$ (pointing to the O atoms) is higher than $d_{x2-y2}$. The half-filling of $d_{xz}$ states in the *P4/mmm* phase gives rise to the metallic antiferromagnetic state (see the SM). However, in the *Pmmn* phase with the dimerizations of V-V atoms, two $d_{xz}$ of the V-V dimer form the occupied bonding state (being the A1@2b EBR) and the unoccupied anti-bonding state.

In summary, we have performed a serious measurements of high-pressure x-ray diffraction, Hall coefficient, magnetoresistance and resistivity on the single crystal $KV_2Se_2O$ up to 60 GPa. Within the studied pressure range, two phases of *P4/mmm* and *Pmmn* can be clearly seen. In the initial phase (0-22 GPa), the metallic behavior maintains this region. The $T_{SDW}$ decreases monotonically to lower temperatures and extrapolates to zero at a critical pressure of around 10 GPa. The pressure range for *Pmmn* phase is from 15 to 60 GPa. The metallic state maintains the low pressures from 15 to near 35 GPa. Upon further compression, the insulating state appears gradually, and dominates the entire phase diagram above 50 GPa, depending on the extrapolated phase boundary. The extracted gap Δ first increases with increasing pressure, and then decreases continuously upon further compression after passing the maximum value ~40 meV at around 43.5 GPa. Obviously, the MI transition should not be induced by structural transition, but the electronic structure near $E_F$. The evolution of MR with pressure and the carrier-type transition indicates the large change of the Fermi surface. The calculations with collinear magnetic configurations demonstrate that the MI transition of $KV_2Se_2O$ under pressure comes from a magnetic obstructed atomic insulator, a state described as spin-singlet state with bonding orbital order.

**I. Methods**

**Sample synthesis.** The polycrystalline $KV_2Se_2O$ used in the high-pressure XRD was synthesized through a solid-state reaction method. The detailed synthetic procedure for this high-quality sample is provided in a previous report [29]. The single crystals were synthesized using KSe as a flux. A mixture of precursors corresponding to K:V:Se:O = 6:2:7:1 was loaded into a Nb container. This vessel was enclosed in an evacuated quartz tube positioned inside an alumina crucible. The reaction was heated to 1000°C for 20 h, followed by controlled cooling (2°C/h) to 650°C and subsequent furnace cooling. The obtained black crystals display a metallic luster and require inert atmosphere storage due to their reactivity with air and humidity.

**Ambient-pressure characterizations.** At ambient pressure, the single-crystal structure of $KV_2Se_2O$ was characterized by using the single crystal x-ray diffractometer system (Micro-Max007HF Mo model) made by Rigaku. The wavelength of Mo K$\alpha$ radiation is 0.719 Å with the tube voltage kV and tube current mA. The system has been calibrated by using $CeO_2$ as the standard material. The temperature dependent resistivity ($\rho_{xx}$) at ambient pressure was measured by commercial cryostat by a Keithley 6221 current source and a 2182A nanovoltmete with a standard four-probe method. The typical size of the sample in this step was 11.1 × 3.1 × 0.1 $mm^3$. The links between the platinum wire and the sample were AB conducting adhesive. The measured temperature range for this step is from 4.5 to 300 K.

**Synchrotron powder X-ray diffraction measurements.** At high pressures, the synchrotron X-ray diffraction patterns were collected at BL15U1, Shanghai Synchrotron Radiation Facility (SSRF, China). The wavelength was chosen as 0.6199 Å. Here, the high pressures of this study were generated by a symmetrical diamond anvil cell (DAC) with the 300 μm culets. The pressure transmitting medium in the sample chamber was the silicon oil. A ruby ball is loaded to serve as internal pressure standard [41]. The obtained two-dimensional XRD patterns were integrated into the one-dimensional function with the help of the DIOPTAS software [42]. Then, the data was further analyzed by using the software of Jana based on the Rietveld method [43].

**High-pressure electrical transports measurements.** The temperature dependent resistivity

$\rho_{xx}$ at high pressures for Run 1 were measured by a commercial cryostat (Janis Research) from 4.5 to 300 K with a Keithley 6221 current source and a 2182 A nanovoltmeter. The measurements of high-pressure $\rho_{xx}$ from 1.7 to 300 K and Hall resistivity $\rho_{xy}$ for Run 2 were performed in another commercial cryostat equipped with a 9 T magnet. The high-pressure electrical measurements were realized by two nonmagnetic DACs with anvils of 300 μm culets. The rhenium gaskets, insulated by the mixture of cubic boron nitride and epoxy, were drilled to create sample chambers approximately 120 μm in diameter. The sizes of the samples were about 80×40×5μm³ and 40×40×5μm³ in Run 1 and Run 2, respectively. In order to maintain the hydrostatic pressure environment, the ammonia borane was loaded in the sample chamber as pressure-transmitting medium. Small ruby balls were loaded into the sample chamber with samples for calibrating the pressure [41].

**Theoretical calculations.** Structure search based on a layer-by-layer slip reconstruction mechanism with the cell and atomic positions optimized under a wide pressure range of 5–60 GPa was performed. This search resulted in the identification of $KV_2Se_2O$ in a tetragonal structure with the space group of *Pmmn*. The structure was then relaxed by *ab* initio calculations under pressure, followed by checks of its relative stability in energy and confirmation of its dynamical stability by phonon mode analysis. First-principles calculations were performed within the framework of density functional theory (DFT) using the projector augmented-wave (PAW) method, as implemented in the Vienna ab initio Simulation Package (VASP) [44,45]. A 6 × 6 × 8 Monkhorst-Pack *k*-point mesh was used, and the plane-wave energy cutoff was set to 480 eV. The LDA+U correction was applied with U = 2 eV. Phonon spectra were obtained using the finite-displacement method on a 2 × 2 × 1 supercell, as implemented in the Phonopy package [46,47]. The surface state was calculated using the WannierTools package [48].

**Reference:**

1. Imada, M., Fujimori, A. & Tokura, Y. Metal-insulator transitions. *Rev. Mod. Phys.* **70**, 1039

(1998).

2. Gebhard, F. The Mott metal-insulator transition: Models and methods. *Springer Berlin Heidelberg* **137**, 1-48 (2000).

3. Abrahams, E. & Gabriel K. The metal-insulator transition in correlated disordered systems. *Science* **274**, 1853-1854 (1996).

4. Ao, L. P., Rice, T. M. & Anderson. P. W. Fluctuation effects at a Peierls transition. *Phys. Rev. Lett.* **31**, 462 (1973).

5. Barbara, M. *et al,* Evidence for a Peierls phase-transition in a three-dimensional multiple charge-density waves solid. *Proc. Natl. Acad. Sci.* **109**, 5603-5608 (2012).

6. Per, B. & Pokrovsky, V. L. Theory of metal-insulator transition in Peierls systems with nearly half-filled bands. *Phys. Rev. Lett.* **47**, 958 (1981).

7. Brandow, B. H. Electronic structure of Mott insulators. *Adv. Phys.* **26**, 651-808 (1977).

8. Gao, S. *et al,* Discovery of a single-band Mott insulator in a van der waals flat-band compound. *Phys. Rev. X* **13**, 041049 (2023).

9. Thouless, D. J. Anderson's theory of localized states. *J. Phys. C: Solid State Phys.* **3**, 1559 (1970).

10. Braganca, H. *et al,* Anderson localization effects near the Mott metal-insulator transition. *Phys. Rev. B* **92,** 125143 (2015).

11. Phillips, P. W., Yeo, L. & Huang, E. W. Exact theory for superconductivity in a doped Mott insulator. *Nat. Phys.* **16**, 1175–1180 (2020).

12. Lee, P. A., Nagaosa, N. & Wen, X. G. Doping a Mott insulator: Physics of high-temperature superconductivity. *Rev. Mod. Phys.* **78**, 17 (2006).

13. Austin, I. G. & Mott, N. F. Metallic and nonmetallic behavior in transition metal oxides. *Science* **168**, 71 (1970).

14. Zhu, M. *et al.* Colossal magnetoresistance in a Mott insulator via magnetic field-driven insulator-metal transition. *Phys. Rev. Lett.* **116**, 216401 (2016).

15. Dillemans, L. Evidence of the metal-insulator transition in ultrathin unstrained $V_2O_3$ thin films. *Appl. Phys. Lett.* **104**, 071902 (2014).

16. High-temperature ferrimagnetic order triggered metal-to-insulator transition in $CaCu_3Ni_2Os_2O_{12}$. *Nat. Commun*. **16**, 3746 (2025).

17. Chernyshov, D. *et al*. Pressure-induced insulator-to-metal transition in $TbBaCo_2O_{5.48}$. *Phy. Rev. Lett*. **103**, 125501 (2009).

18. Laukhin, V., Fontcuberta, J., Garcia-Munoz, J. L. & Obradors. X. Pressure effects on the metal-insulator transition in magnetoresistive manganese perovskites. *Phy. Rev. B* **56**, R10009 (1997).

19. Greenberg, E. *et al*. Pressure-induced site-selective Mott insulator-metal transition in $Fe_2O_3$. *Phys. Rev. X* **8**, 031059 (2018).

20. Gavriliuk, A. G., Trojan, I. A. & Struzhkin, V. V. Insulator-metal transition in highly compressed NiO. *Phys. Rev. Lett*. **109**, 086402 (2012).

21. Tafti, F. F., Ishikawa, J. J., McCollam, A., Nakatsuji, S. & Julian, S. R. Pressure-tuned insulator to metal transition in $Eu_2Ir_2O_7$. *Phys. Rev. B* **85**, 205104 (2012).

22. Matsuoka, T. *et al*. Pressure-induced insulator-to-metal transition in the van der Waals compound $CoPS_3$. *Phys. Rev. B* **107**, 165125 (2023).

23. Chen, L. C. *et al*. Valence fluctuation driven superconductivity in orthorhombic lead telluride. *Phys. Rev. B* **05**, 174503 (2022).

24. Modak, P. & Verma, A. K. Pressure induced multi-centre bonding and metal–insulator transition in $PtAl_2$. *Phys. Chem. Chem. Phys*. **21**, 13337 (2019).

25. Matsuoka, T. & Shimizu, K. Direct observation of a pressure-induced metal-to-semiconductor transition in lithium. *Nature* **458**, 186 (2009).

26. Irizawa, A. *et al*. Direct observation of a pressure-induced metal-insulator transition in $LiV_2O_4$ by optical studies. *Phys. Rev. B* **84**, 235116 (2011).

27. Song, C. *et al*. Altermagnets as a new class of functional materials. *Nat. Rev. Mater*. **321**, 554-557 (2008).

28. Jiang, B. *et al*. A metallic room-temperature *d*-wave altermagnet. *Nat. Phys*. **21**, 754 (2025).

29. Bai J. L. *et al*. Absence of long-range order in the vanadium oxychalcogenide $KV_2Se_2O$

with nontrivial band topology. *Phys. Rev. B* **110**, 165151 (2024).

30. Zhuang, H. Y. *et al*. Charge transfer caused anomalies of physical properties of $KV_2Se_2O$. *EPL* **150**, 36003 (2025).

31. Moon, E. G. & Sachdev, S. Competition between spin density wave order and superconductivity in the underdoped cuprates. *Phys. Rev. B* **80**, 035117 (2009).

32. Chen, H. *et al*. Coexistence of the spin-density wave and superconductivity in $Ba_{1-x}K_xFe_2As_2$. *Europhys. Lett.* **85**, 17006 (2009).

33. Zhang, J. J. *et al*. Intertwined density waves in a metallic nickelate. *Nat. Commun*. **11**, 6003 (2020).

34. Kusmartseva, A. F., Sipos, B., Berger, H., Forro, L. & Tutis, E. Pressure induced superconductivity in Pristine *1T*-$TiSe_2$. *Phys. Rev. Lett.* **103**, 236401 (2009).

35. Chi, Z. *et al*. Superconductivity in pristine $2H_a$-$MoS_2$ at ultrahigh pressure. *Phys. Rev. Lett.* **120**, 037002 (2018).

36. Petricek, V., Dusek, M. & Palatinus, L. Crystallographic computing system JANA2006: General features. *Z. Kristallogr. Cryst. Mater.* **229**, 345 (2014).

37. Das, L. *et al*. Two-carrier magnetoresistance: Applications to $Ca_3Ru_2O_7$. *J. Phys. Soc. Jpn.* **90**, 054702 (2021).

38. Ma, K. Y. *et al*. Correlation between the dome-shaped superconducting phase diagram, charge order, and normal-state electronic properties in $LaRu_3Si_2$. *Nat. Commun*. **16**, 6149 (2025).

39. Gao, J. C., Wu, Q. S., Persson, Clas. & Wang, Z. J. Irvsp: To obtain irreducible representations of electronic states in the VASP. *Comput. Phys. Commun.* **216**, 107760 (2021).

40. Gao, J. C. *et al*. Unconventional materials: the mismatch between electronic charge centers and atomic positions. *Sci. Bull.* **67**, 598–608 (2022).

41. Mao, H. K., Xu, J. A. & Bell, P. M. Calibration of the ruby pressure gauge to 800 kbar under quasi-hydrostatic conditions. *J. Geophys. Res. Solid Earth* **91**, 4673-4676 (1986).

42. Prescher, C. & Prakapenka, V. B. DIOPTAS: a program for reduction of two-dimensional

X-ray diffraction data and data exploration. *High Press. Res.* **35,** 223-230 (2015).

43. Petricek, V., Dusek, M. & Palatinus, L. Crystallographic computing system JANA2006: General features. *Z. Kristallogr.* **229**, 345-352 (2014).

44. Kresse, G. & Furthmüller, Efficient iterative schemes for ab initio total-energy calculations using a plane-wave basis set. J. *Phys. Rev. B* **54**, 11169 (1996).

45. Perdew, J. P., Burke, K. & Ernzerhof, M. Generalized gradient approximation made simple. *Phys. Rev. Lett.* **77**, 3865 (1996).

46. Togo, A. First-principles phonon calculations with phonopy and phono3py. *J. Phys. Soc. Jpn.* **92**, 012001 (2023).

47. Togo, A. *et al.* Implementation strategies in phonopy and phono3py. *J. Phys. Condens. Matter* **35**, 353001 (2023).

48. Wu, Q. S. *et al*. Wanniertools: An open-source software package for novel topological materials. *Comput. Phys. Commun.* **224**, 405–416 (2018).

**Acknowledgments** The work was supported by the National Key Research and Development Program of China (Grant Nos. 2024YFA1611300, 2023YFA1608900, 2021YFA1400300, 2020YFA0711502), the National Natural Science Foundation of China (Grant Nos.12404163, 12375304, 92263202, 12374020), the Strategic Priority Research Program of the Chinese Academy of Sciences (Grant No. XDB33000000). We thank the high-pressure synergetic measurement station (A9) of Synergtic Extreme Condition User Facility (SECUF) for the help in high-pressure resistivity and single-crystal x-ray diffraction measurements, and the beamline 15U1 at the Shanghai Synchrotron Radiation Facility (SSRF) for the high-pressure XRD measurements.

**Author Contributions** X.Y. designed and coordinated this project. J.C. and G.C. synthesized the samples. L.C. performed the electrical transport and high-pressure X-ray diffraction measurements. Z.Z. helps to collected the single-crystal XRD data at ambient pressure. J.W. searched the high-pressure structures. J.Y. and Z.W. carried out the ab-initio calculations and theoretical analyses. L.C., J.Y., Z.W. and X.Y. wrote the paper with the inputs from other authors. The manuscript reflects the contributions of all authors.

**Additional Information** Correspondence and requests for materials should be addressed to G.F.C., J.T.W., Z.J.W. and X.H.Y. (gfchen@iphy.ac.cn, wjt@iphy.ac.cn, wzj@iphy.ac.cn, yuxh@iphy.ac.cn).

**Competing financial interests** The authors declare no competing financial interests.

**Figures and captions:**

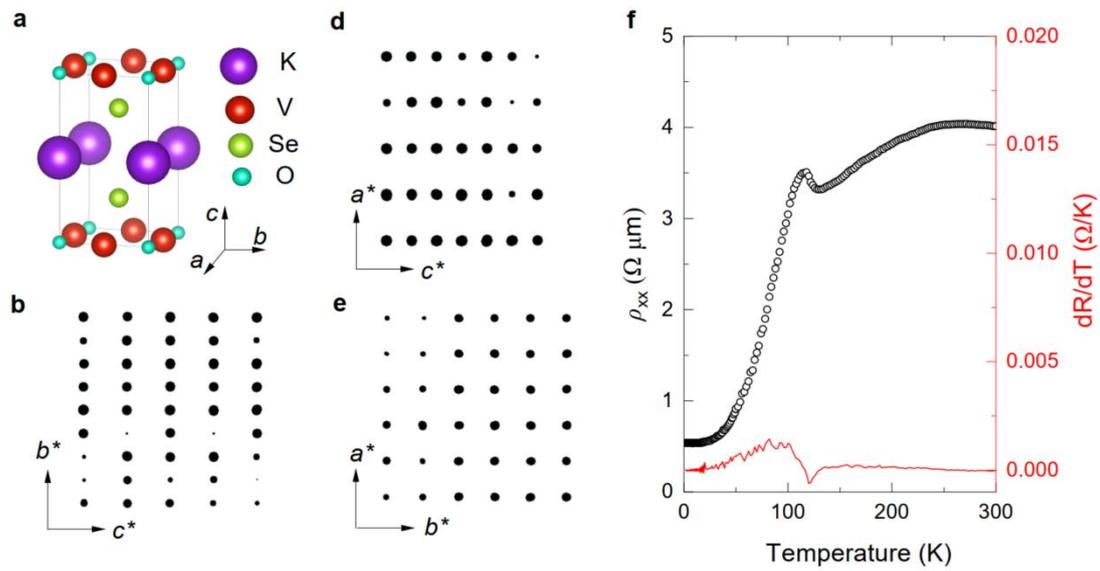

**Figure 1 Single-crystal x-ray diffraction data and resistivity of $KV_2Se_2O$ at ambient pressure. a** Schematic crystal structures of $P4/mmm$ phase at low pressures. **b-e** Reciprocal lattice data of $KV_2Se_2O$ viewed along $a^*$, $b^*$, and $c^*$ axis, respectively. The size of the spots represents the intensity of the diffraction peaks. **f** Temperature dependence of resistivity $\rho_{xx}$ and dR/dT at ambient pressure. The valley in dR/dT presents the transition temperature of SDW.

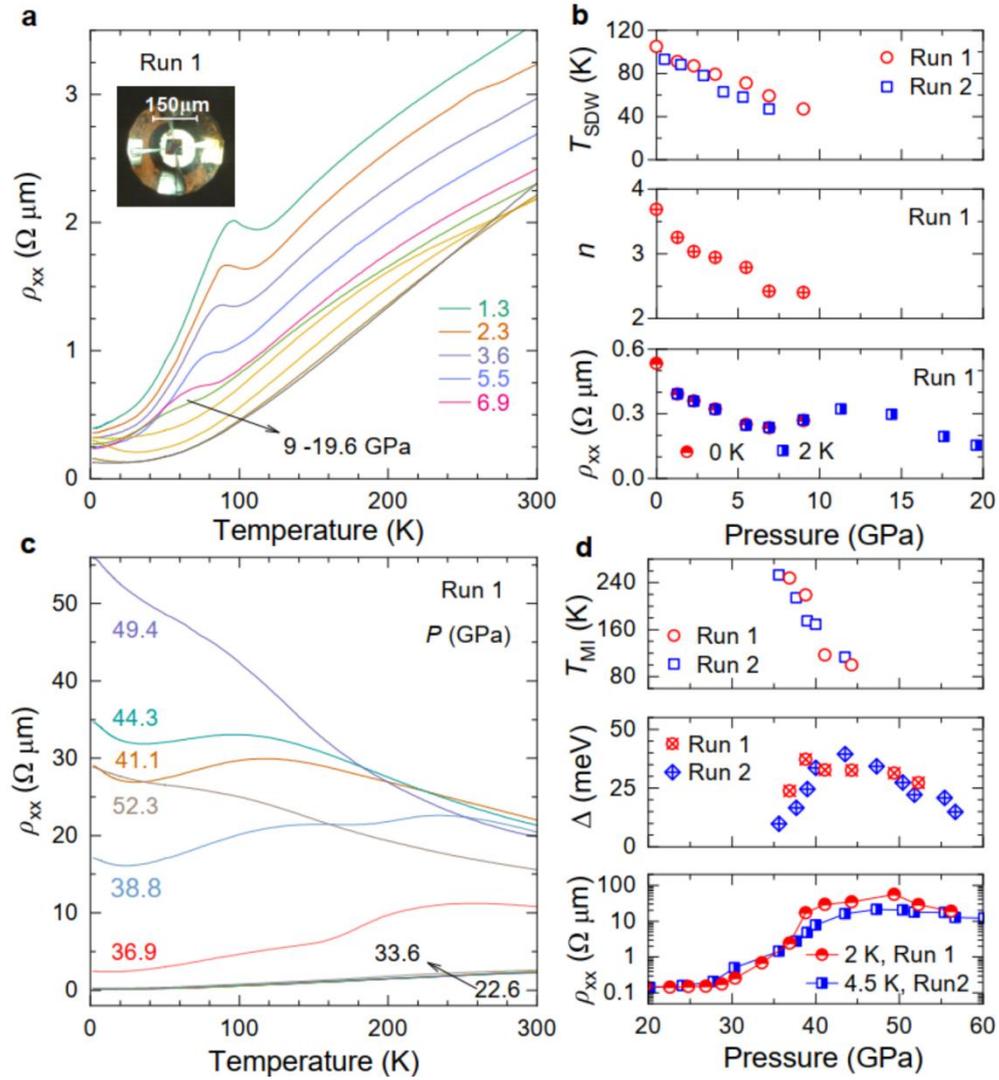

**Figure 2 Spin density wave and metal to insulator transition of $KV_2Se_2O$ under pressure.**
**a** Temperature dependent resistivity at high pressures from 1.3 to 19.6 GPa in Run 1. The insets show the actual set-up for the electrical transport measurements under pressure in the first run. **b** Evolution of SDW-transition temperature of $KV_2Se_2O$ with pressure in Run 1 and Run 2, respectively (upper panel). The pressure dependence of the parameter $n$ (middle panel). The dots in the lower panel represent the extracted resistivity $\rho_{xx}$ at 0 and 2 K, respectively. **c** Temperature dependence of resistivity at various pressures range from 22.6 to 52.3 GPa in the first run. **d** Pressure dependent the transition temperature of metal-to-insulator $T_{MI}$ (upper panel) in two separated runs (Run1 and Run 2). The middle penal shows the pressure-dependent band-gap $\Delta$ in the two runs. The lower penal displays the resistivity $\rho_{xx}$ at 2 K in run 1 and 4.5 K in run 2, respectively.

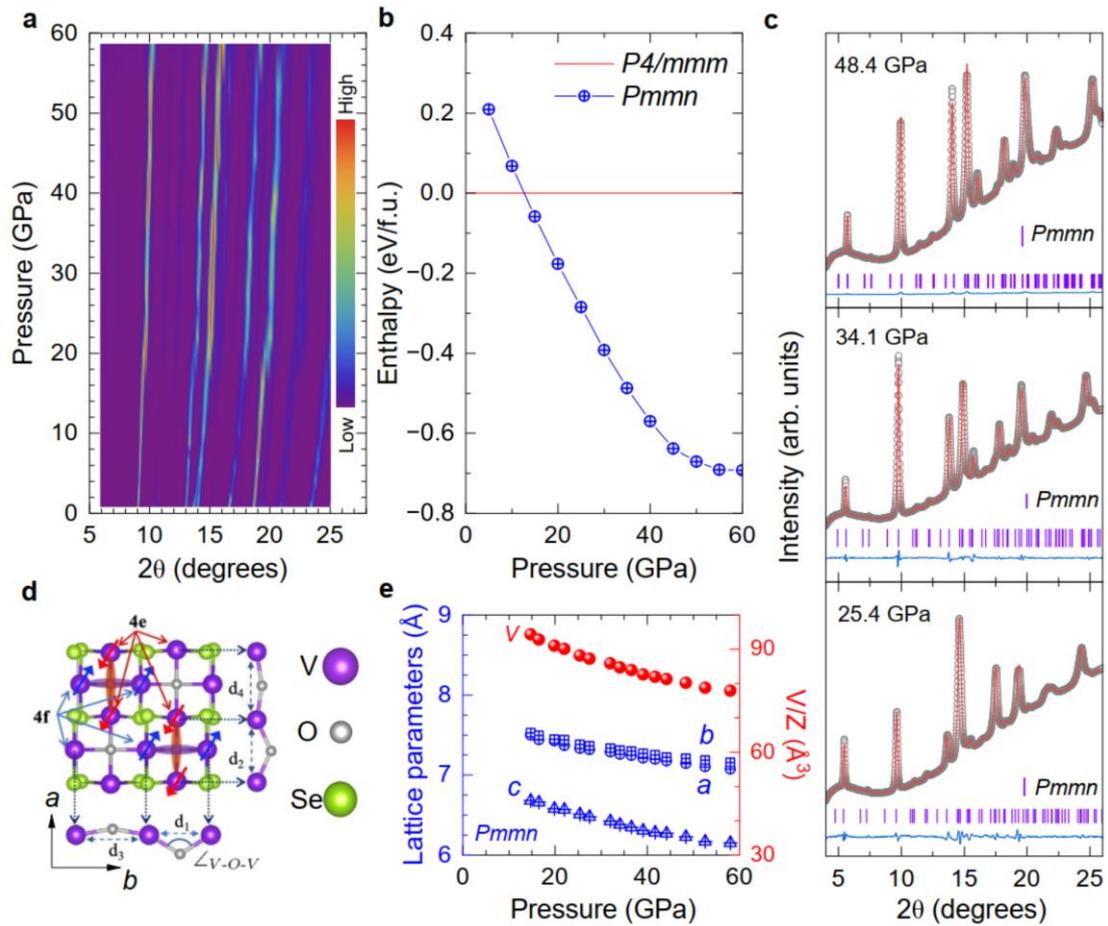

**Figure 3 The crystal structures of KV$_2$Se$_2$O under pressure. a** Synchrotron XRD patterns of the sample at room temperature and various pressures up to 58 GPa. The red and blue colors represent the high and low diffraction intensity, respectively. **b** Enthalpy of the P4mm phase as a function of pressure. **c** Representative Rietveld refinements of the XRD patterns at 25.4, 34.1 and 48.4 GPa, respectively. The open circles are the experimental data points. The red curves are the calculated results based on the different structures. The thin curves at the bottom of each panel are the deviations between experiments and calculations. **d** Crystal structure and of KV$_2$Se$_2$O viewed from *c* axis, d represents the V-V distance. **e** Pressure-dependent lattice parameter(s) and unit-cell volume in difference phases, respectively.

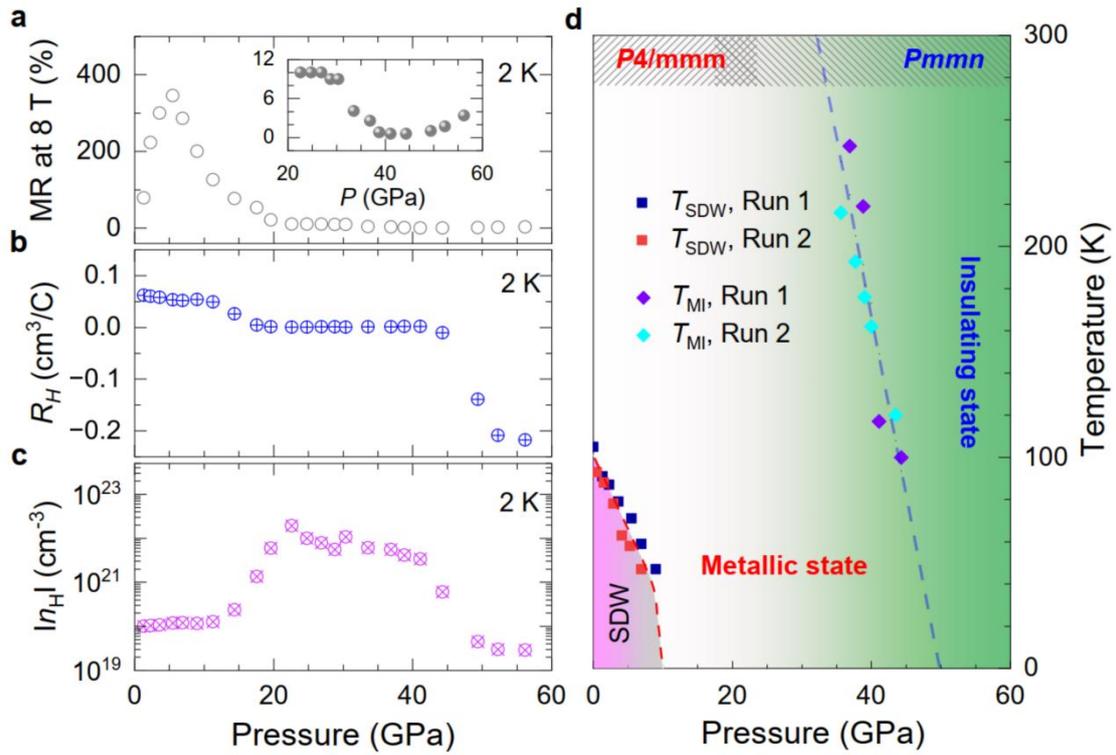

**Figure 4 Pressure dependence of magnetoresistance MR, Hall coefficient $R_H$ and carrier concentration $n_H$ and the temperature-pressure (T-P) phase diagram of of $KV_2Se_2O$. a** Evolution of MR at 8 T measured at 2 K with pressure. The insert shows the MR between 20 and 56 GPa. **b** Pressure dependence of the $R_H$ extracted by a liner fitting of Hall resistivity $\rho_{xy}$ at 2 K. **c** The evolution of the absolute value of $n_H$ with pressure extracted by a semiclassical one-band model. **d** Phase diagram of $KV_2Se_2O$ determined by the data of resistivity and XRD. The shaded area at the top represents the pressure range of the *P4/mmm* and *Pmmn* phases. The overlapping region between 15 and 22 GPa means the mixed phase. At low pressures, P<10 GPa, the squares represent the transition temperature of SDW. The red dashed line is the boundary of SDW guided by eyes. The rhombus at higher pressures means the temperature of MI transition, $T_{MI}$. The linear extrapolation between 35 and 50 GPa is the boundary of $T_{MI}$ guided by eyes.

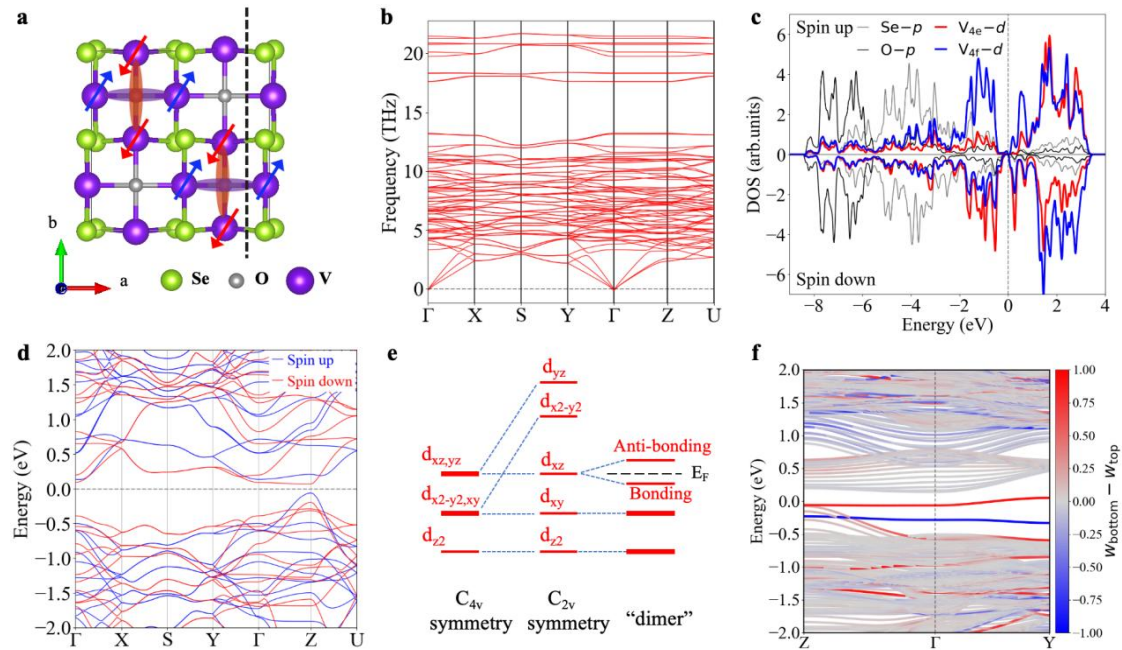

**Figure 5 DFT results of *Pmmn* phase KV$_2$Se$_2$O at 43.5 GPa. a** Crystal structure of *Pmmn* phase, distorted from the 2×2 supercell of *P*4/mmm phase. The spin-up (spin-down) V are the 4f (4e) Wycyoff positions. **b** Phonon spectrum with no negative-frequency modes. **c** Partial density of states of Se-*p*, O-*p*, V$_{4e}$-*d* and V$_{4f}$-*d* orbitals. **d** Spin-polarized LSDA+U band structure. A clear band gap is shown, in sharp contrast to the metallic one of *P*4/mmm phase. **e** Schematic diagram of the orbital evolution under the crystal field in this compound. This shows the formation of the insulating behavior of the $d^{2.5}$ configuration of V$^{+2.5}$ state. **f** (100)-surface spectrum, cutting through the dashed line in **a**. The red-colored (left-surface) and blue-colored (right-surface) bands in the gap of the bulk projection are the surface states.